\newcommand{\be}{\begin{equation}}
\newcommand{\bea}{\begin{eqnarray}}
\newcommand{\ee}{\end{equation}}
\newcommand{\eea}{\end{eqnarray}}
\newcommand{\nn}{\nonumber}

\newcommand{\qa}{\alpha}
\newcommand{\qb}{\beta}

\newcommand{\qG}{\Gamma}
\newcommand{\qd}{\delta}

\newcommand{\qe}{\varepsilon}

\newcommand{\qy}{\theta}

\newcommand{\qk}{\kappa}

\newcommand{\qL}{\Lambda}

\newcommand{\qs}{\sigma}

\newcommand{\qt}{\tau}
\newcommand{\qf}{\varphi}
\newcommand{\qF}{\Phi}

\newcommand{\qJ}{\Psi}
\newcommand{\qo}{\omega}
\newcommand{\qO}{\Omega}
\newcommand{\sgn}{{\rm sgn}}

\newcommand{\Erfc}{{\rm Erfc}}

\renewcommand{\Im}{{\rm Im}\,}
\newcommand{\rd}{{\rm d}}

\newcommand{\half}{\mbox{$\textstyle \frac{1}{2}$}}

\newcommand{\EE}{{\mathbb E}}

\newcommand{\NN}{{\mathbb N}}

\newcommand{\PP}{{\mathbb P}}

\newcommand{\cC}{{\mathcal C}}

\newcommand{\cO}{{\mathcal O}}

\newcommand{\cQ}{{\mathcal Q}}

\newcommand{\vecp}{{\boldsymbol{p}}}
\newcommand{\vecv}{{\boldsymbol{v}}}

\newcommand{\vecx}{{\boldsymbol{x}}}
\newcommand{\vecz}{{\boldsymbol{z}}}
\newcommand{\vecsig}{{\boldsymbol{\qs}}}
\newcommand{\vecone}{{\mathbf 1}}
\newcommand{\pr}{{\rm Pr}}

\documentclass[10pt]{llncs}
\usepackage{amsfonts}
\usepackage{amsmath}
\usepackage{amssymb}
\usepackage{url}
\usepackage{epsfig}
\usepackage{graphics}
\usepackage{graphicx}

\begin{document}

\setlength{\parindent}{0mm}

\title{Asymptotically false-positive-maximizing attack on non-binary Tardos codes}
\author{Antonino Simone and Boris \v{S}kori\'{c}}

\date{}
\institute{Eindhoven University of Technology}

\maketitle

\begin{abstract}
We use a method recently introduced by Simone and \v{S}kori\'{c} 
to study accusation probabilities for non-binary Tardos fingerprinting codes.
We generalize the pre-computation steps in this approach to include a broad 
class of collusion attack strategies.
We analytically derive properties of a special attack that 
asymptotically maximizes false accusation probabilities.
We present numerical results on sufficient code lengths for this attack,
and explain the abrupt transitions that occur in these results.
\end{abstract}
%
%=================== Introduction ========================
\section{Introduction}
%
%----------------- Forensic watermarking --------------------
\underline{\bf 1.1 Collusion attacks against forensic watermarking}.

Watermarking provides a means for tracing the origin and
distribution of digital data. Before distribution of digital content,
the content is modified by applying an imperceptible watermark (WM), 
embedded using a watermarking algorithm. 
Once an unauthorized copy
of the content is found, it is possible to trace those users who
participated in its creation. 
This process is known as `forensic watermarking'.
Reliable tracing requires resilience against attacks that aim
to remove the WM.  Collusion attacks,
where a group of pirates cooperate, are a
particular threat: differences between their versions of the content
tell them where the WM is located.
Coding theory has produced a number of
collusion-resistant codes. The resulting system has two
layers~\cite{hw06tifs,Schaathun08}: 
The coding layer determines which message to embed and protects against collusion attacks. The
underlying watermarking layer hides symbols of the code in segments
of the content.
The interface between the layers is usually specified in terms of the \textit{Marking
Assumption} plus additional assumptions that are referred to as a
`model'. The Marking Assumption states that the colluders are able to
perform modifications only in those segments where they received different WMs.  
These segments are called detectable positions. The `model' specifies the kind of symbol
manipulations that the attackers are able to perform {\em in detectable
positions}. 
In the \textit{Restricted Digit Model} (RDM) the attackers must choose one of the symbols
that they have received.
The \textit{unreadable digit model}
also allows for erasures.  
In the
\textit{arbitrary digit model} the attackers can choose arbitrary symbols, while the \textit{general digit model}
additionally allows erasures.

%----------------- Tardos codes -------------------------
\underline{\bf 1.2 Tardos codes}.

Many collusion resistant codes have been proposed in the literature. 
Most notable are the
Boneh-Shaw construction~\cite{BS1998} and the by now famous 
Tardos code~\cite{Tardos}. The former uses a concatenation of
an inner code with a random outer code, while the latter one is a
fully randomized binary code. 
In Tardos' original paper \cite{Tardos} a binary code
was given achieving length
$m=100c_0^2\lceil\ln\frac{1}{\qe_1}\rceil$, 
along with a proof that $m\propto c_0^2$ is
asympotically optimal
 for large coalitions, for all alphabet sizes. 
Here $c_0$ denotes the number of colluders
to be resisted, and $\qe_1$ is the maximum allowed probability
of accusing a fixed innocent user. 
Tardos' original construction had two unfortunate design choices which caused the high
proportionality constant 100. 
(i) The false negative probability $\qe_2$ (not accusing any attacker)
was set as $\qe_2=\qe_1^{c_0/4}$, even though
$\qe_2\ll\qe_1$ is highly unusual in the context of content distribution;
a deterring effect is achieved already at $\qe_2\approx\half$, while $\qe_1$ needs to be very small.
In the subsequent literature (e.g. \cite{SVCT,BlayerTassa}) the $\qe_2$ was decoupled from $\qe_1$,
substantially reducing $m$.
(ii) The symbols 0 and 1 were not treated equally.
Only segments where the attackers produce a 1 were taken into account.
This ignores 50\% of all information.
A fully symbol-symmetric version of the scheme was given in \cite{symmetric},
leading to a further improvement of $m$ by a factor~4.
A further improvement was achieved in \cite{Nuida}. 
The code construction contains a step where a bias parameter is randomly set for each segment.
In Tardos' original construction the probability density function (pdf) for the
bias is a continuous function.
In \cite{Nuida} a class of discrete distributions was given that performs better than the original pdf
against finite coalition sizes.
In \cite{XFF,CombinedDigit} the Marking Assumption was relaxed, and the accusation algorithm 
of the nonbinary Tardos code was modified to effectively cope with signal processing attacks such as averaging and
addition of noise.

All the above mentioned work followed the so-called `simple decoder' approach, i.e. an accusation score is
computed for each user, and if it exceeds a certain threshold, he is considered suspicious.
One can also use a `joint decoder' which computes scores for sets of users.
Amiri and Tardos \cite{AmiriTardos} have given a capacity-achieving
joint decoder construction for the binary code.
(Capacity refers to the information-theoretic treatment \cite{Merhav,Moulin08,Huang09} of the attack as
a channel.) 
However, the construction is rather impractical, requiring computations for many candidate coalitions.
In~\cite{symmetric} the binary construction was generalized to $q$-ary alphabets, 
in the simple decoder approach.
In the RDM,
the transition to a larger alphabet size has benefits beyond the mere fact that a $q$-ary symbol
carries $\log_2 q$ bits of information.

%------------------- Gaussian approx -------------------------------------
\underline{\bf 1.3 The Gaussian approximation}.

The Gaussian approximation, introduced in \cite{SVCT}, is
a useful tool in the analysis of Tardos codes.
The assumption is that the accusations are normal-distributed.
The analysis is then drastically simplified;
in the RDM
the scheme's performance is almost completely determined by a single parameter, 
the average accusation $\tilde\mu$ of the coalition (per segment).
The sufficient code length against a coalition of size $c$ is
$m=(2/\tilde\mu^2)c^2 \ln(1/\qe_1)$.
The Gaussian assumption is motivated by the Central Limit Theorem (CLT):
An accusation score consists of a sum of i.i.d. per-segment contributions.
When many of these get added, the result is close to normal-distributed: 
the pdf is close to Gaussian in a region around the average, 
and deviates from Gaussian in the tails.
The larger $m$ is, the wider this central region.
In \cite{SVCT,symmetric} it was argued
that in many practical cases the central region is sufficiently wide
to allow for application of the Gaussian approximation.
In \cite{TardosFourier} a semi-analytical
method was developed for determining the exact shape
of the pdf of innocent users' accusations, without simulations. 
This is especially useful in the case of very low accusation probabilities,
where simulations would be very time-consuming.
The false accusation probabilities
were studied for two attacks: majority voting and interleaving.

%--------------------------- Contributions ---------------------------
\underline{\bf 1.4 Contributions}.

We discuss the simple decoder in the RDM, choosing $\qe_2\approx\half$.
We follow the approach of \cite{TardosFourier} for computing false accusation
probabilities.
Our contribution is threefold:

1. 
We prove a number of theorems (Theorems \ref{th:class1}--\ref{th:class3})
that allow efficient computation of
pdfs for more general attacks than the ones treated in \cite{TardosFourier}.

2. 
We identify which attack minimizes the all-important\footnote{
Asymptotically for large $m$, the $\tilde\mu$-minimizing attack is the `worst case' attack in the RDM
in the sense that the false accusation probability is maximized.
} 
parameter $\tilde\mu$.
It was shown in \cite{TardosFourier} that the majority voting attack achieves this for
certain parameter settings, but we consider more general parameter values.
We derive some basic properties of the attack.

3.
We present numerical results for the $\tilde\mu$-minimizing attack.
When the coalition is small the graphs contain sharp transitions; 
we explain these transitions as an effect of the abrupt changes in pdf shape  
when the attack turns from majority voting into minority voting.

%=================== Preliminaries ========================
\section{Notation and preliminaries}
\label{sec:prelim}

We briefly describe the $q$-ary version of the Tardos code as introduced in \cite{symmetric}
and the method of \cite{TardosFourier} to compute innocent accusation probabilities.

%------------------- q-ary Tardos -------------------------------------
\underline{\bf 2.1 The $q$-ary Tardos code}.

The number of symbols in a codeword is $m$.
The number of users is $n$.
The alphabet is $\cQ$, with size~$q$.
$X_{ji}\in\cQ$ stands for the $i$'th symbol in the codeword of user~$j$.
The whole matrix of codewords is denoted as~$X$.

\underline{\it Two-step code generation}.
$m$ vectors $\vecp^{(i)}\in [0,1]^q$ are independently drawn
according to a distribution $F$, with
\be
	F(\vecp)=\qd(1-\sum_{\qb\in\cQ}p_\qb)\cdot\frac1{B(\qk\vecone_q)}\prod_{\qa\in\cQ}p_{\qa}^{-1+\qk}.
\label{defF}
\ee
Here $\vecone_q$ stands for the vector $(1,\cdots,1)$ of length $q$, 
$\qd(\cdot)$ is the Dirac delta function,
and $B$ is the generalized Beta function. 
$\qk$ is a positive constant. 
%For $q=2$ it is optimal to set $\qk=1/2$.
For $v_1,\cdots,v_n>0$
the Beta function is defined as\footnote{
This is also known as a Dirichlet integral.
The ordinary Beta function ($n=2$) is $B(x,y)=\qG(x)\qG(y)/\qG(x+y)$.
}
\be
	B(\vecv)=\int_0^1\!\rd x^n\; \qd(1-\sum_{a=1}^n x_a)\prod_{b=1}^n x_b^{-1+v_b}
	=\frac{\prod_{a=1}^n \qG(v_a)}{\qG(\sum_{b=1}^n v_b)}.
\label{defBeta}
\ee
All elements $X_{ji}$ are drawn independently
according to $\pr[X_{ji}=\qa | \vecp^{(i)}] = p_\qa^{(i)}$.

\underline{\it Attack}.
The coalition is $\cC$, with size $c$.
%The part of $X$ observed by the coalition is~$X_\cC$.
The $i$'th segment of the pirated content contains  
a symbol $y_i\in\cQ$. 
We define vectors $\vecsig^{(i)}\in\NN^q$ as
\be
	\qs^{(i)}_\qa \;\triangleq\;  |\{ j\in\cC: X_{ji}=\qa  \}|
\label{defqs}
\ee
satisfying $\sum_{\qa\in\cQ}\qs^{(i)}_\qa=c$.
In words: $\qs_\qa^{(i)}$ counts how many colluders have received symbol $\qa$ in segment~$i$.
The attack strategy may be probabilistic.
As usual,
it is assumed that this strategy is column-symmetric, symbol-symmetric and attacker-symmetric.
It is expressed as probabilities $\qy_{y|\vecsig}$ 
that apply independently for each segment. Omitting the column index,
\be
	\Pr[y|\vecsig]= \qy_{y|\vecsig}.
\label{defqy}
\ee

\underline{\it Accusation}.
The watermark detector sees the symbols $y_i$.
For each user $j$, the {\em accusation sum} $S_j$ is computed,
\bea
	S_j=\sum_{i=1}^m S_j^{(i)}
	&\quad\mbox{where}\quad&
	S_j^{(i)}=\;\; g_{[X_{ji}==y_i]}(p_{y_i}^{(i)}),
\label{Sj}
\eea
where the expression $[X_{ji}==y_i]$ evaluates to 1 if $X_{ji}=y_i$ and to 0 otherwise,
and the functions $g_0$ and $g_1$ are defined as
\bea
	g_1(p)  \triangleq  \sqrt{\frac{1-p}{p}}
	\quad&;&\quad
	g_0(p)  \triangleq  -\sqrt{\frac{p}{1-p}}.
\label{g0g1}
\eea
The total accusation of the coalition is
$S:=\sum_{j\in\cC} S_j$.
The choice (\ref{g0g1}) is the unique choice that satisfies
\bea
	pg_1(p)+(1-p)g_0(p)=0
	&\quad ; \quad &
	p [g_1(p)]^2+(1-p) [g_0(p)]^2=1.
\label{gproperties}
\eea
This has been shown to have optimal properties for $q=2$ \cite{Furon08,SVCT}.
Its unique properties (\ref{gproperties}) also hold for $q\geq 3$; that is the main motivation
for using (\ref{g0g1}).
A user is `accused' if his accusation sum exceeds a threshold~$Z$, i.e. $S_j>Z$. 

The parameter $\tilde\mu$ is defined as $\frac1m \EE[S]$, where $\EE$ stands for the expectation
value over all random variables.
The $\tilde\mu$ depends on $q$, $\qk$, the collusion strategy, and weakly on $c$.
In the limit of large $c$ it converges to a finite value, and the code length scales as
$c^2/\tilde\mu^2$.

%------------------- prelim marginals -------------------------------------
\underline{\bf 2.2 Marginal distributions and strategy parametrization}.

Because of the independence between segments, the segment index will be dropped from
this point onward.
For given $\vecp$, the vector $\vecsig$ is multinomial-distributed, 
$\PP(\vecsig|\vecp)={c\choose\vecsig}\prod_\qa p_\qa^{\qs_\qa}$.
Averaged over $\vecp$, the $\vecsig$ has distribution
$\PP(\vecsig)={c\choose\vecsig}
\frac{B(\qk\vecone_q+\vecsig)}{B(\qk\vecone_q)}$. 
Two important marginals were given in \cite{TardosFourier}.
First, 
the marginal probability 
$\PP_1(b)\triangleq \pr[\qs_\qa=b]$ for one arbitrary component~$\qa$,
\be
	\PP_1(b)=
	{c\choose b}\frac{B(\qk+b, \qk[q-1]+c-b)}{B(\qk,\qk[q-1])}.
\ee
Second, given that $\qs_\qa=b$, the probability that the remaining $q-1$ components of the vector $\vecsig$
are given by $\vecx$,
\be
	\PP_{q-1}(\vecx|b)=
%	\pr[\vecsig_{\setminus\qa}=\vecx| \qs_\qa=b]=
	{c-b\choose\vecx}
	\frac{B(\qk\vecone_{q-1}+\vecx)}{B(\qk\vecone_{q-1})}.
\label{defPqmin1}
\ee
It is always implicit that $\sum_{\qb\in\cQ\setminus\{\qa\}}x_\qb=c-b$.

An alternative parametrization was introduced for the collusion strategy, which
exploits the fact that 
(i) $\qy_{\qa|\vecsig}$ is invariant under permutation of the symbols
$\neq\qa$; 
(ii) $\qy_{\qa|\vecsig}$ depends on $\qa$ only through the value of $\qs_\qa$.
 
\be
	\qJ_b(\vecx) \triangleq \qy_{\qa|\vecsig}
	\mbox{ given that }
	\qs_\qa=b \mbox{ and }\vecx=\mbox{the other components of }\vecsig.
\label{defPsi}
\ee
Thus, $\qJ_b(\vecx)$ is the probability that the pirates choose a symbol that they have seen $b$ times,
given that the other symbols' occurences are $\vecx$.
Strategy-dependent parameters $K_b$ were introduced as follows,
\be
	K_b \triangleq 
	\EE_{\vecx|b} \qJ_b(\vecx)=
	\sum_\vecx \PP_{q-1}(\vecx|b)
	\qJ_b(\vecx).
\label{defK}
\ee
Due to the marking assumption $K_0=0$ and $K_c=1$.
The $K_b$ obey the sum rule $q\sum_{b=0}^c K_b \PP_1(b)=1$.
Efficient
pre-computation of the $K_b$ parameters can speed up the computation of
a number of quantities of interest, among which the $\tilde\mu$ parameter.
It was shown that $\tilde\mu$ can be expressed as
\be
	\tilde\mu=\sum_\vecsig \PP(\vecsig)\sum_{\qa\in\cQ}\qy_{\qa|\vecsig}T(\qs_\qa)
	=
	q\sum_{b=0}^c K_b \PP_1(b)T(b),
\label{mutilde}
\ee
where
\be
	T(b)\triangleq \left\{ \half-\qk+\frac{b}{c}(\qk q-1) \right\}
	c\frac{\qG(b+\qk-\half)}{\qG(b+\qk)}
	\frac{\qG(c-b+\qk[q-1]-\half)}{\qG(c-b+\qk[q-1])}.
\label{defT}
\ee

%------------------- prelim pdf -------------------------------------
\underline{\bf 2.3 Method for computing false accusation probabilities}.

The method of \cite{TardosFourier} is based on the convolution rule for generating functions
(Fourier transforms): Let $A_1\sim f_1$ and $A_2\sim f_2$ be continuous random variables, and let $\tilde f_1$, $\tilde f_2$
be the Fourier transforms of the respective pdfs.
Let $A=A_1+A_2$. Then the easiest way to compute the pdf of $A$ (say $\qF$) is to use the fact that
$\tilde \qF(k)=\tilde f_1(k) \tilde f_2(k)$.
If $m$ i.i.d. variables $A_i\sim\qf$ are added, $A=\sum_i A_i$, then the pdf of $A$
is found using $\tilde \qF(k)=[\tilde\qf(k)]^m$.
In \cite{TardosFourier} the pdf $\qf$ was derived for an innocent user's one-segment accusation
$S_j^{(i)}$. The Fourier transform was found to be
\be
	\tilde\qf(k) = 
	\frac{2q}{B(\qk,\qk[q-1])}
	\sum_{b=1}^c {c\choose b}K_b\cdot\left[
	\qL(d_b, v_b; k)+\qL(v_b-1, d_b+1; -k)
	\vphantom{\int}\right],
\label{phitilde}
\ee
with 
\[
	d_b \triangleq b+\qk \quad ; \quad
	v_b \triangleq c-b+\qk[q-1]+1
\]
\[
\qL(d,v; k) =
	(-ik)^{2v} \qG(-2v)
	\; {}_1 F_2(v+d; v+\half, v+1; \frac{k^2}4)
	+ \half\sum_{j=0}^\infty \frac{(ik)^j}{j!}B(d+\frac j2, v-\frac j2).
\]
Using this result for $\tilde\qf$ it is then possible to cast the expression $\tilde\qf^m$
in the following special form,
\be
	\left[\tilde\qf(\frac k{\sqrt m})\right]^m=e^{-\half k^2}\left[1+
	\sum_{t=0}^\infty \qo_t(m) (i\,\sgn\,k)^{\qa_t}|k|^{\nu_t}
	\right],
\label{coeffsnuchipi}
\ee
where $\qa_t$ are real numbers;
the coefficients $\qo_t(m)$ are real;
the powers $\nu_t$ satisfy $\nu_0>2$, $\nu_{t+1}>\nu_t$.
In general the $\nu_t$ are not all integer.
The $\qo_t$ decrease with increasing $m$ as $m^{-\nu_t/6}$ or faster.
Computing all the $\qa_t$, $\qo_t$, $\nu_t$ up to a certain cutoff $t=t_{\rm max}$
is straightforward but laborious, and leads to huge expressions if done analytically;
it is best done numerically, e.g. using series operations in Mathematica.
Once all these coefficients are known, the false accusation probability 
is computed as follows.
Let $R_m$ be a function defined as
$R_m(\tilde Z):=\pr[S_j >\tilde Z \sqrt m]$
(for innocent $j$). Let $\qO$ be the corresponding function in case the pdf of $S_j$ is Gaussian,
$\qO(\tilde Z)=\half\Erfc(\tilde Z/\sqrt 2)$.
Then
\be
\label{probcorrections}
	R_m(\tilde Z) = \qO(\tilde Z)+
	\frac{1}{\pi}\sum_{t=0}^\infty \qo_t(m)
	\qG(\nu_t)2^{\nu_t/2}\Im\left[
	i^{-\qa_t}H_{-\nu_t}(i\tilde Z/\sqrt2)\right].
\ee
Here $H$ is the Hermite function. It holds that $\lim_{m\to\infty}R_m(\tilde Z)=\qO(\tilde Z)$.
For a good numerical approximation it suffices to take terms up to some cutoff $t_{\rm max}$.
The required $t_{\rm max}$ is a decreasing function of $m$.

%=================== Results ========================
\section{Our results}
\label{sec:results}

%------------------- Kb theorems -------------------------------------
\underline{\bf 3.1 Computing $K_b$ for several classes of colluder strategy}.

Our first contribution is a prescription for efficiently computing the
$K_b$ parameters for more general colluder strategies than those studied in \cite{TardosFourier}.
We consider the strategy parametrization $\qJ_b(\vecx)$ with $b\neq 0$.
The vector $\vecx\in\NN^{q-1}$ can contain several entries equal to $b$. 
The number of such entries will be denoted as~$\ell$. 
(The dependence of $\ell$ on $b$ and $\vecx$ is suppressed in the notation for the sake of brevity.)
The number of remaining entries is $r\triangleq q-1-\ell$.
These entries will be denoted as $\vecz=(z_1,\cdots,z_r)$, with $z_j\neq b$ by definition.
Any strategy possessing the symmetries mentioned in Section~\ref{sec:prelim}
can be parametrized as a function $\qJ_b(\vecx)$ which in turn can be expressed
as a function of $b$, $\ell$ and $\vecz$; it is invariant under permutation of the entries in $\vecz$.
We will concentrate on the following `factorizable' classes of attack, each one a sub-class of the previous one.
\begin{description}
\item[Class 1:]
$\qJ_b(\vecx)$ is of the form $w(b,\ell)\prod_{k=1}^r W(b,\ell,z_k)$
\item[Class 2:]
$\qJ_b(\vecx)$ is of the form $\frac{w(b)}{\ell+1} \prod_{k=1}^r W(b,z_k)$
\item[Class 3:]
$\qJ_b(\vecx)$ is of the form $\frac1{\ell+1}\prod_{k=1}^r W(b,z_k)$,
with $W(b,z_k)\in\{0,1\}$
and $W(b,z_k)+W(z_k,b)=1$. By definition $W(b,0)=1$.
\end{description}

Class~1 merely restricts the dependence on $\vecz$ to a form factorizable in the components $z_k$.
This is a very broad class, and contains e.g. the interleaving attack 
($\qy_{\qa|\vecsig}=\frac{\qs_\qa}c$, $\qJ_b(\vecx)=\frac bc$) which has no dependence on $\vecz$.

Class~2 puts a further restriction on the $\ell$-dependence.
The factor $1/(\ell+1)$ implies that symbols with equal occurrence have equal probability of
being selected by the colluders. (There are $\ell+1$ symbols that occur $b$ times.)

Class~3 restricts the function $W$ to a binary `comparison' of its two arguments:
$\qJ_b(\vecx)$ is nonzero only if $b$ is `better' than $z_k$
for all $k$, i.e. $W(b,z_k)=1$.
An example of such a strategy is majority voting, where $\qJ_b(\vecx)=0$ if there exists a $k$
such that $z_k>b$, and $\qJ_b(\vecx)=\frac1{\ell+1}$ if $z_k<b$ for all $k$.
Class~3 also contains minority voting, and in fact any strategy
which uses a strict ordering or `ranking' of the occurrence counters $b$, $z_k$.
(Here a zero always counts as `worse' than nonzero.)

Our motivation for introducing classes 1 and 2 is mainly technical, 
since they affect to which extent the $K_b$ parameters can be computed analytically.
In the next section we will see that
class~3 captures not only majority/minority voting but also the $\tilde\mu$-reducing attack.

\begin{theorem}
\label{th:class1}
Let $N_b\in\NN$ satisfy $N_b>\max\{c-b,|c-bq|,(c-b)(q-2)\}$. 
Let $\qt_b\triangleq e^{i2\pi/N_b}$, and let
\bea
	G_{ba\ell}\triangleq \!\!\!\!\!\!
	\sum_{z\in\{0,\ldots,c-b\}\setminus\{b\}}
	\!\!\!\!\!\!\!\!\!\!\!\!
	\frac{\qG(\qk+z)W(b,\ell,z)}{\qt_b^{az}z!},
	& \quad\quad &
	v_{ba}\triangleq\frac{\qG(\qk+b)}{\qt_b^{ab}b!}.
\label{defGv}
\eea
Then for strategies in class~1 it holds that
\[
	K_b= \frac{(c-b)!}{N_b\qG(c-b+\qk[q-1])B(\qk\vecone_{q-1})}
	\sum_{a=0}^{N_b-1}\qt_b^{a(c-b)}\sum_{\ell=0}^{q-1}
	\binom{q-1}{\ell} G_{ba\ell}^{q-1-\ell}w(b,\ell) v_{ba}^\ell. 
\]
\end{theorem}

\begin{theorem}
\label{th:class2}
For strategies in class~2 the quantity $G_{ba\ell}$ 
as defined in (\ref{defGv}) does not depend on $\ell$ and can be denoted as
$G_{ba}$ (with $W(b,\ell,z)$ replaced by $W(b,z)$). It then holds that
\[
	K_b = \frac{b!(c-b)! \; w(b)}{q N_b \qG(\qk+b) 
	\qG(c-b+\qk[q-1])B(\qk\vecone_{q-1})}
	\sum_{a=0}^{N_b-1}\qt_b^{ac}
	\left[\left( G_{ba}+v_{ba}\right)^q -G_{ba}^q \right].
\]
\end{theorem}

\begin{theorem}
\label{th:class3}
For strategies in class~3, Theorem~\ref{th:class2} holds, where 
$w(b)=1$ and
$G_{ba}$ can be expressed as
\be
	G_{ba}=\!\!\!\!\!\!
	\sum_{\stackrel{ z\in\{0,\ldots,c-b\}\setminus\{b\} }{W(b,z)=1}}
	\!\!\!\!\!\!
	\frac{\qG(\qk+z)}{\qt_b^{az}z!}.
\label{Gba3}
\ee
\end{theorem}
The proofs of Theorems \ref{th:class1}--\ref{th:class3} are given in the Appendix.
Without these theorems, straightforward computation of $K_b$ following (\ref{defK})
would require a full sum over $\vecx$, which for large $c$ comprises
$\cO(c^{q-2}/(q-1)!)$ different terms. 
($q-1$ variables $\leq c-b$, with one constraint, and with permutation symmetry.
We neglect the dependence on~$b$.)
Theorem~\ref{th:class1} reduces the number of terms to $\cO(q^2 c^2)$ at worst;
a factor $c$ from computing $G_{ba}$, a factor $q$ from $\sum_\ell$
and a factor $N_b$ from $\sum_a$, with $N_b<qc$.
In Theorem~\ref{th:class2} the $\ell$-sum is eliminated, resulting in $\cO(qc^2)$ terms.

We conclude that, for $q\geq5$ and large $c$,
 Theorems \ref{th:class1} and \ref{th:class2} can significantly reduce the
time required to compute the $K_b$ parameters.\footnote{
To get some feeling for the orders of magnitude:
The crossover point where $qc^2=c^{q-2}/(q-1)!$
lies at $c=120$, 27, 18, 15, 13, for $q=$5, 6, 7, 8, 9
respectively.
}
A further reduction occurs in Class~3 if the $W(b,z)$ function is zero for many~$z$.

\vskip2mm

%------------------- Identifying the attack -------------------------------------
\underline{\bf 3.2 The $\tilde\mu$-minimizing attack}.

%Our second contribution is pointing out which $\qJ_b(\vecx)$
%corresponds to the `asymptotically worst case' attack.
Asymptotically for large code lengths 
the colluder strategy has negligible impact on the Gaussian shape of the innocent
(and guilty) accusation pdf.
For $q\geq 3$
the main impact of their strategy is on the value of the statistical parameter $\tilde\mu$.
(For the binary symmetric scheme with $\qk=\half$, the $\tilde\mu$ is fixed
at $\frac2\pi$; the attackers cannot change it. 
Then the strategy's impact on the pdf shape is {\em not}
negligible.)

Hence for $q\geq 3$ the strategy that minimizes $\tilde\mu$ is asymptotically a
`worst-case' attack in the sense of maximizing the false positive probability.
This was already argued in \cite{symmetric}, and it was shown
how the attackers can minimize $\tilde\mu$.
From the first expression in (\ref{mutilde}) it is evident that, for a given $\vecsig$,
the attackers must choose the symbol $y$ such that $T(\qs_y)$ is minimal, i.e.
$y=\arg \min_\qa T(\qs_\qa)$.
In case of a tie it does not matter which of the best symbols is chosen, and without
loss of generality we impose symbol symmetry, i.e. if the minimum $T(\qs_\qa)$
is shared by $N$ different symbols, then each of these symbols will have probability $1/N$
of being elected.
Note that this strategy fits in class~3.
The function $W(b,z_k)$ evaluates to 1 if $T(b)<T(z_k)$ and to 0 otherwise.\footnote{
For $x,y\in\NN$, with $x\neq y$,
it does not occur in general that $T(x)=T(y)$.
The only way to make this happen is to choose $\qk$ in a very special way
as a function of $q$ and~$c$.
W.l.o.g.
we assume that $\qk$ is not such a pathological case.
}

Let us introduce the notation $x=b/c$, $x\in(0,1)$.
Then for large $c$ we have \cite{TardosFourier}
\be
	T(cx)\approx \frac{\half-\qk+x(\qk q-1)}{\sqrt{x(1-x)}}.
\label{Tapprox}
\ee

From (\ref{Tapprox}) we deduce some elementary properties of the function $T$.
\begin{itemize}
\item
If $\qk<\frac1{2(q-1)}$ then $T$ is monotonically decreasing, and
$T(b)$ may become negative at large $b$.
\item
If $\qk>\half$, then $T$ is monotonically
increasing, and $T(b)$ may become negative at small $b$.
\item
For $\qk$ in between those values, $T(b)$ is nonnegative and has a minimum at
$\frac bc\approx \frac1{q-2}(\frac1{2\qk}-1)$.
\end{itemize}
We expect that the existence of negative $T(b)$ values has a very bad impact
on $\tilde\mu$, and hence that $\qk$ is best chosen in the interval
$(\frac1{2(q-1)},\half)$.

\begin{figure}[t]
\begin{center}
\includegraphics[width=6cm]{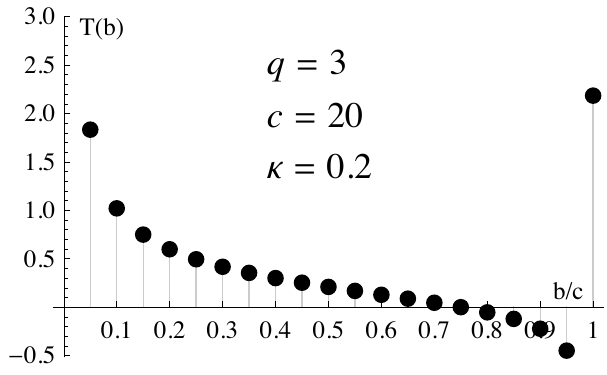}
\includegraphics[width=6cm]{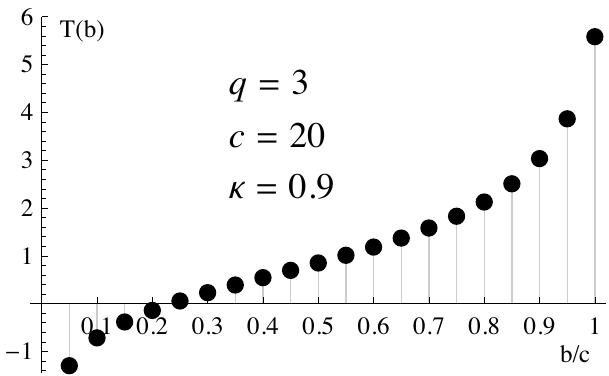}
\caption{\it The function $T(b)$ for $q=3$, $c=20$ and  two values $\qk$
outside $(\frac1{2[q-1]},\half)$.
}
\label{fig:Tb}
\end{center}
\end{figure}

Fig.~\ref{fig:Tb} shows the function $T(b)$ for two values of $\qk$ outside
this `safe' interval.
For $\qk=0.2$ it is indeed the case that $T(b)<0$ at large $b$,
and for $\qk=0.9$ at small~$b$. 
Note that $T(c)$ is always positive due to the Marking Assumption.
For small $\qk$, the $T(b)$-ranking of the points is clearly such that
majority voting is the best strategy; similarly, for large $\qk$ minority voting is best.
For intermediate values of $\qk$ a more complicated ranking will occur.

\vskip2mm
%------------------- Numerics -------------------------------------
\underline{\bf 3.3 Numerical results for the $\tilde\mu$-minimizing attack}.

In \cite{TardosFourier} the $\tilde\mu$-minimizing attack was studied 
for a restricted parameter range, $\qk\approx 1/q$.
For such a choice of $\qk$ the strategy reduces to majority voting.
We study a {\em broader range}, applying the full $\tilde\mu$-minimizing attack.
We use Theorem~\ref{th:class3} to precompute the $K_b$ and then 
(\ref{phitilde}),
(\ref{coeffsnuchipi}) and (\ref{probcorrections}) to compute the false accusation probability~$R_m$
as a function of the accusation threshold.
We found that keeping terms in the expansion with $\nu_t\leq 37$ gave stable results.

For a comparison with \cite{TardosFourier}, we set 
$\qe_1=10^{-10}$, and search for the smallest codelength $m_*$ for which it holds that 
$R_m(\tilde\mu \sqrt m/c)\leq\qe_1$. The special choice $\tilde Z=\tilde\mu \sqrt m/c$
puts the threshold at the expectation value of a colluder's accusation.
As a result the probability of a false negative error is $\approx\half$.
Our results for $m_*$ are consistent with the numbers given in \cite{TardosFourier}.

\begin{figure}
\includegraphics[width=6cm]{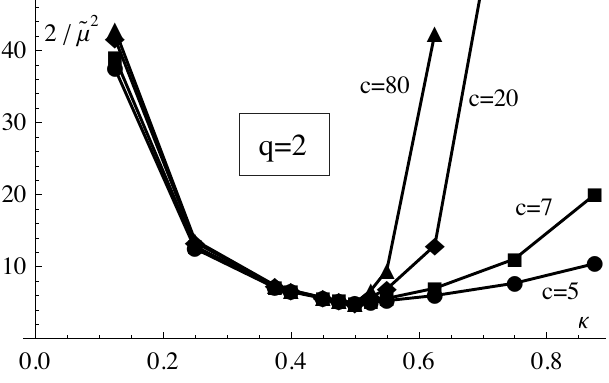}
\includegraphics[width=6cm]{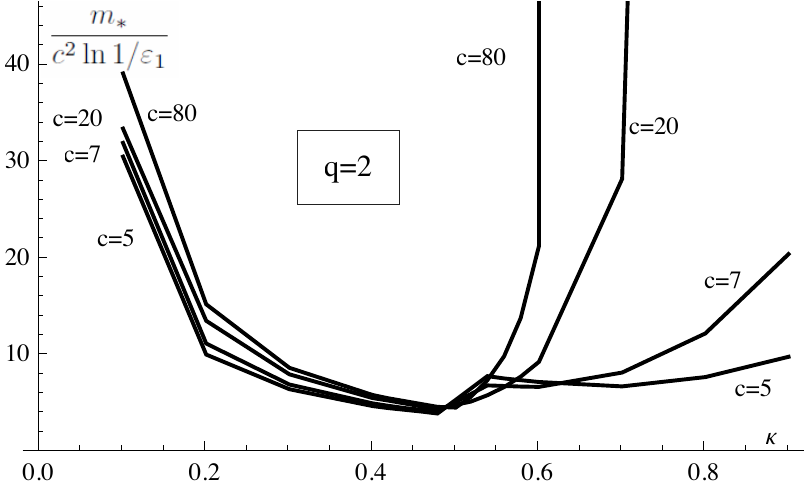}
\\
\includegraphics[width=6cm]{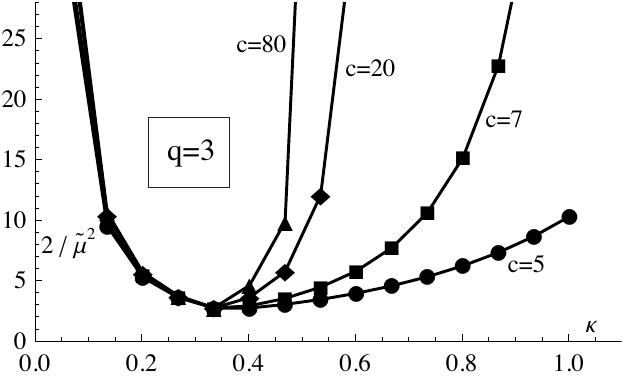}
\includegraphics[width=6cm]{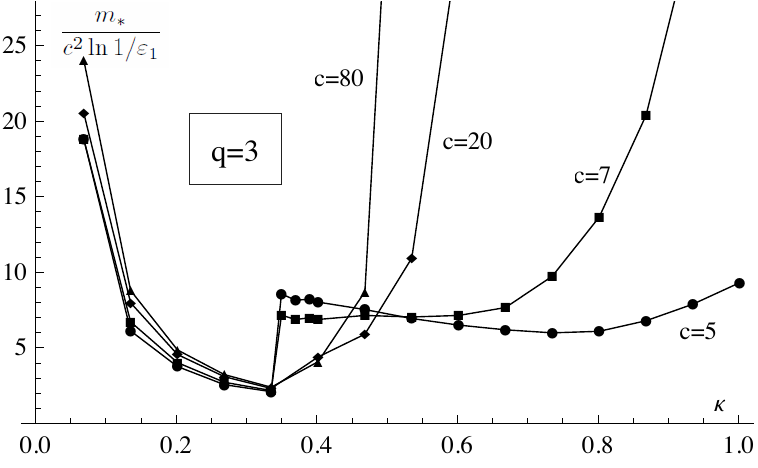}
\\
\includegraphics[width=6cm]{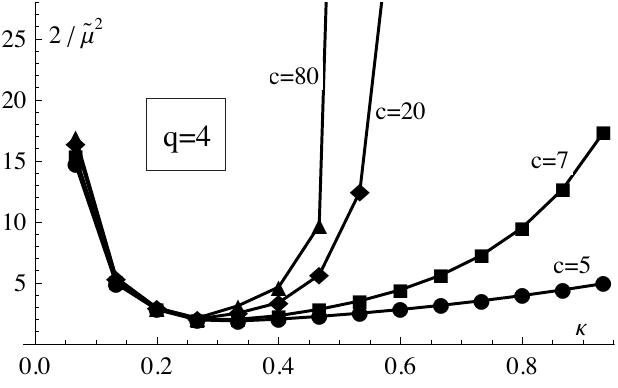}
\includegraphics[width=6cm]{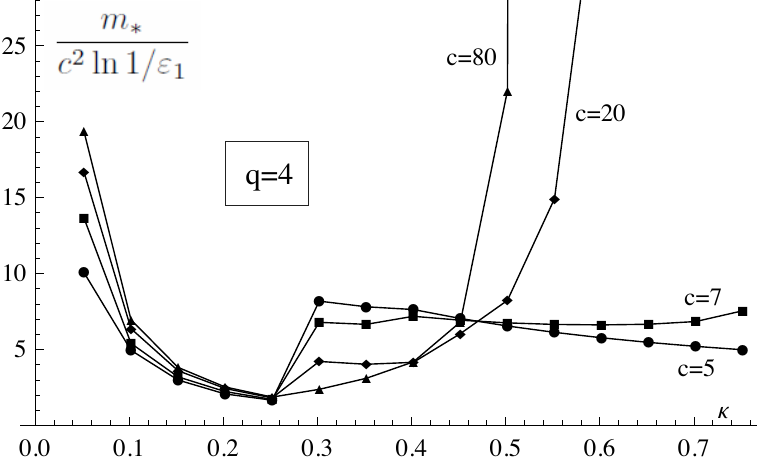}
\\
\includegraphics[width=6cm]{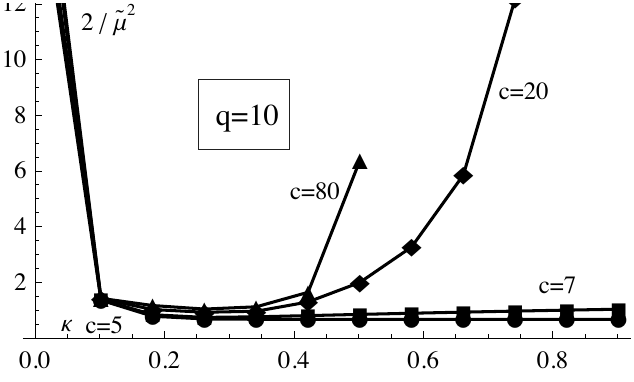}
\includegraphics[width=6cm]{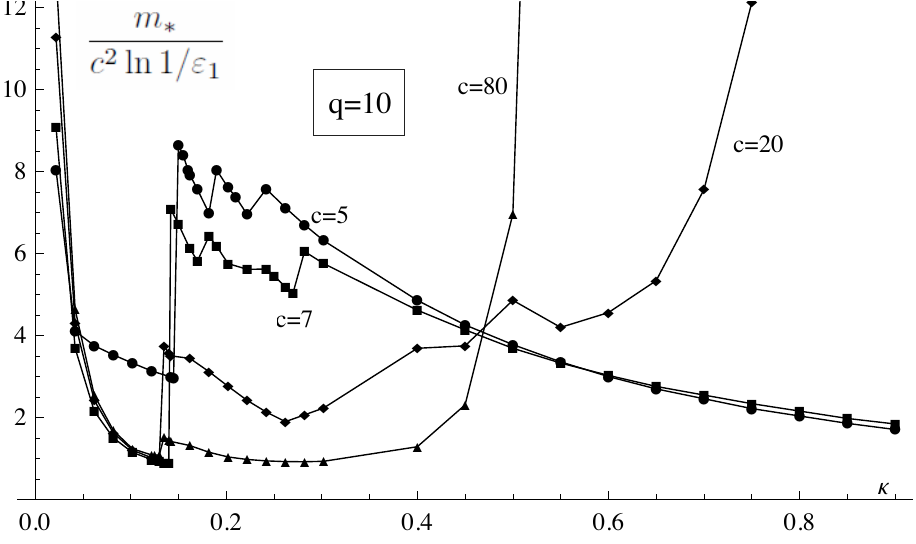}
\caption{\it 
Numerical results for the $\tilde\mu$-minimizing attack. $\qe_1=10^{-10}$.
{\bf Left:} The Gaussian-limit code length constant $\frac2{\tilde\mu^2}$ as a function of $\qk$,
for various $q$
and $c$.
{\bf Right:} The sufficient code length $m_*$, scaled by the factor $c^2\ln(1/\qe_1)$
for easy comparison to the Gaussian limit.
}
\label{fig:m}
\end{figure}

In Fig.~\ref{fig:m}
we present graphs of $2/\tilde\mu^2$ as a function of $\qk$ for various $q$, $c$.
\footnote{
The $\tilde\mu$ can become negative.
These points are not plotted, as they represent
a situation where the accusation scheme totally fails, and there exists no sufficient code length $m_*$.)
}
If the accusation pdf is Gaussian, then
the quantity $2/\tilde\mu^2$ is very close to the proportionality constant in the equation
$m\propto c^2\ln(1/\qe_1)$.
We also plot $\frac {m_*}{c^2\ln(1/\qe_1)}$ as a function of $\qk$
for various $q$, $c$.
Any discrepancy between the $\tilde\mu$ and $m_*$ plots is caused by non-Gaussian tail shapes.

In the plots on the left we see that 
the attack becomes very powerful (very large $2/\tilde\mu^2$) around $\qk=\half$, especially
for large coalitions. This can be understood from the fact that the $T(b)$ values are decreasing,
and some even becoming negative for $\qk>\half$,
as discussed in Section~3.2.
This effect becomes weaker when $q$ increases.
The plots also show a strong deterioration of the scheme's performance when
$\qk$ approaches $\frac1{2(q-1)}$, as expected.

For small and large $\qk$,
the left and right graphs show roughly the same behaviour.
In the middle of the $\qk$-range, however, the $m_*$ is very irregular.
We think that this is caused by rapid changes in the `ranking' of $b$ values
induced by the function $T(b)$; there is a transition from majority voting 
(at small $\qk$) to minority voting (at large $\qk$).
It was shown in \cite{TardosFourier} that 
(i) majority voting causes a more Gaussian tail shape than minority voting;
(ii) increasing $\qk$ makes the tail more Gaussian.
These two effects together explain the $m_*$ graphs in Fig.~\ref{fig:m}:
first, the transition for majority voting to minority voting
makes the tail less Gaussian (hence increasing $m_*$), and then increasing $\qk$
gradually makes the tail more Gaussian again (reducing $m_*$).

%\clearpage

\begin{figure}
\begin{center}
\includegraphics[width=10cm]{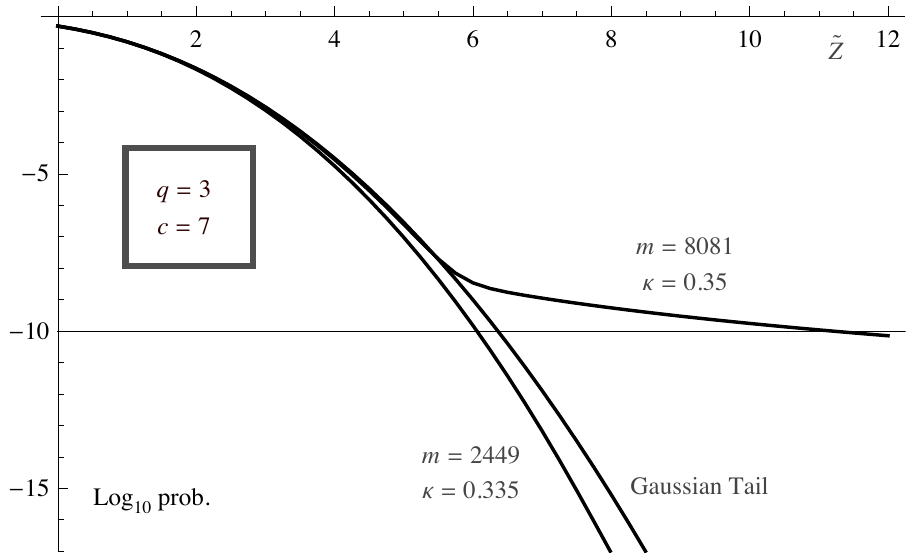}
\caption{\it
Accusation probability for a fixed innocent user as a function of the 
(scaled) accusation threshold $\tilde Z=Z/\sqrt m$.
The attack is the $\tilde\mu$-minimizing attack.
The graph shows the Gaussian limit, and two parameter settings which correspond
to `before' and `after' a sharp transition. 
}
\label{fig:FP}
\end{center}
\end{figure}

In Fig.~\ref{fig:FP} we show the shape of the false accusation pdf 
of both sides of the transition in the $q=3$, $c=7$ plot.
For the smaller $\qk$ the curve is better than Gaussian up to 
false accusation probabilities of better than $10^{-17}$.
For the larger $\qk$ the curve becomes worse than Gaussian around $10^{-8}$,
which lies significantly above the desired $10^{-10}$.

%\vskip1mm

The transition from majority to minority voting is cleanest for $q=2$,
and was already shown in \cite{symmetric} to lie precisely at $\qk=\half$ for all $c$.
For $q\geq 3$ it depends on $c$ and is less easy to pinpoint.

%=================== Discussion ========================
\section{Discussion}
\label{sec:discussion}

We have tested the pdf computation method of \cite{TardosFourier}
for a large range of parameter values and for the various `rankings'
that are part of the $\tilde\mu$-minimizing attack.
The method has performed well under all these conditions.

Our results reveal the subtle interplay between the average colluder accusation $\tilde\mu$ and
the shape of the pdf of an innocent user's accusation sum.
The sharp transitions that occur in Fig.~\ref{fig:m}
show that there is a $\qk$-range (to the left of the transition) 
where the $\tilde\mu$-reducing attack is not optimal for small coalitions.
It is not yet clear what the optimal attack would be there, but certainly
it has to be an attack that concentrates more on the pdf shape than on~$\tilde\mu$, e.g.
the minority voting or the interleaving attack.

For large coalitions the pdfs are very close to Gaussian.
From the optimum points $m_*$ as a function of $\qk$
we see that it can be advantageous to use an alphabet size $q>2$.
(Even if a non-binary symbol occupies $\log_2 q$
times more space in the content than a binary symbol.)

\vskip2mm

{\bf Acknowledgements}\\
Discussions with Dion Boesten, Jan-Jaap Oosterwijk and Benne de Weger
are gratefully acknowledged.

%
%=========================================================

\bibliographystyle{plain}

\bibliography{tardosprob}

\appendix

%======================== PROOFS ======================
\section{Proofs}

%------------------- Proof class 1 -----------------------
\underline{Proof of Theorem~\ref{th:class1}}

We start from (\ref{defK}), with $\PP_{q-1}$ defined in (\ref{defPqmin1}), 
and reorganize the $\vecx$-sum 
to take the multiplicity $\ell$ into account:
\bea
	\sum_{\vecx}[\cdots] 
	&\to& \sum_{\ell=0}^{\ell_{\rm max}}{q-1\choose\ell}
	\sum_{\vecz \in (\{0,\ldots,c-b\}\setminus\{b\})^r}
	\!\!\!\!\!\!
	\qd_{0,c-b(\ell+1)-\sum_{k=1}^r z_k}[\cdots]
	\nn\\
	&=& 
	\sum_{\ell=0}^{\ell_{\rm max}}{q-1\choose\ell}
	\sum_{z_1 \in \{0,\ldots,c-b\}\setminus\{b\}} \cdots 
	\sum_{z_r \in \{0,\ldots,c-b\}\setminus\{b\}}
	\!\!\!\!\!\!
	\qd_{0,c-b(\ell+1)-\sum_{k=1}^r z_k}[\cdots]
	\nn
\label{xsplitup}
\eea
where $\qd$ is the Kronecker delta, and
$\ell_{\rm max}=\min\{ q-1,\lfloor\frac{c-b}b\rfloor \}$.
The factor ${q-1\choose \ell}$ pops up because the summand in (\ref{defK})
is fully symmetric under permutations of~$\vecx$.
The Kronecker delta takes care of the constraint that the components of $\vecz$ add up to~$c-b-\ell b$.

If $\ell_{\rm max}=\lfloor\frac{c-b}b\rfloor$ and 
the sum over $\ell$ is extended beyond $\ell_{\rm max}$, then all the additional
terms are zero, because the Kronecker delta condition cannot be satisfied.
(The $\sum_k z_k$ would have to become negative.)
Hence we are free to replace the upper summation bound $\ell_{\rm max}$ by $q-1$
without changing the result of the sum.

Next we use a sum representation of the Kronecker $\qd$ as follows,
\be
	\qd_{0,s}=\frac{1}{N_b}\sum_{a=0}^{N_b-1}(e^{i2\pi/N_b})^{as},
\label{Kronecker}
\ee
with $s=c-b(l+1)-\sum_k z_k$.
This is a correct representation only if $N_b$ is larger than the maximum $|s|$ that can occur.
The most positive possible value of $s$ is attained at ($\ell=0$, $\vecz=0$), namely
$s=c-b$.
The most negative value ($s_{\rm neg}$)
is attained when $z_k=c-b$ for all $k$. Since there are $r=q-1-\ell$
components in $\vecz$, we have 
$s_{\rm neg}=\min_{\ell}[c-b(\ell+1)-(q-1-\ell)(c-b)]$. The function is linear in $\ell$, so there are
only two candidates: the extreme values $\ell=0$ and $\ell=q-1$,
which yield $|s_{\rm neg}|=(q-2)(c-b)$ and $|s_{\rm neg}|=|c-bq|$ respectively.
Hence $N_b$ has to be larger than $\max\{c-b,(q-2)(c-b),|c-bq|\}$.

Our expression for $K_b$ now contains sums over $\ell$, $z_k$ and $a$.
We shift the $a$-sum completely to the left. Next we write
\be
	B(\qk\vecone_{q-1}+\vecx) = \frac{[\qG(\qk+b)]^\ell \prod_{k=1}^{q-1-\ell}\qG(\qk+z_k)}
	{\qG(c-b+\qk[q-1])},
\label{appbetax}
\ee
\be
	{c-b\choose \vecx} = \frac{(c-b)!}{[b!]^\ell \prod_{k=1}^{q-1-\ell}z_k!}.
\label{appchoosex}
\ee
All the expressions depending on the $z_k$ variables are fully factorized;
the part of the summand that contains the $z_k$ is given by 
\be
	\prod_{k=1}^{q-1-\ell}\left[\sum_{z_k \in\{0,\ldots,c-b\}\setminus\{b\} }
	\frac{W(b,\ell,z_k)\qG(\qk+z_k)}{z_k! \; \qt_b^{az_k}}\right]
	=(G_{ba\ell})^{q-1-\ell}.
\ee
Theorem \ref{th:class1} follows after some elementary rewriting.
\hfill$\square$

\vskip1mm
%------------------- Proof class 2 -----------------------
\underline{Proof of Theorem~\ref{th:class2}}

We start from $K_b$ as given by Theorem~~\ref{th:class1}.
The $G_{ba\ell}$ becomes $G_{ba}$, so the factor $G_{ba}^{q-1}$
can be moved out of the $\ell$-sum. 
The $w(b,\ell)$ becomes $w(b)/(\ell+1)$ and $w(b)$ can also be moved out of the $\ell$-sum.
The remaining sum is $\sum_{\ell=0}^{q-1}{q-1\choose \ell}\frac1{\ell+1}(v_{ba}/G_{ba})^\ell$
which evaluates to
$[(G_{ba}+v_{ba})^q-G_{ba}^q]G_{ba}^{1-q}/(qv_{ba})$.
Theorem~\ref{th:class2} follows after substituting the definition of $v_{ba}$ and some
rewriting.
\hfill$\square$

\vskip1mm
%------------------- Proof class 3 -----------------------
\underline{Proof of Theorem~\ref{th:class3}}

In (\ref{defGv}) the $W(b,\ell,z)$ becomes $W(b,z)$.
The definition of class~3 specifies that $W(b,z)$ is either 1 or 0.
The result (\ref{Gba3}) trivially follows.
\hfill$\square$

\end{document}